\documentclass[twocolumn,amsmath,amssymb,pre,showpacs]{revtex4}

\usepackage{graphicx}
\usepackage{dcolumn}
\usepackage{bm}
\begin{document}
\bibliographystyle{aip}

\title{Increasing thermal rectification: Effects of long range interactions}

\author{Emmanuel Pereira and Ricardo R. \'Avila}
 \email{emmanuel@fisica.ufmg.br; rravila@fisica.ufmg.br}
\affiliation{Departamento de F\'{\i}sica--Instituto de Ci\^encias Exatas, Universidade Federal de Minas Gerais, CP 702,
30.161-970 Belo Horizonte MG, Brazil}

\date{\today}

\begin{abstract}
In this paper, we study the effects of the interparticle interaction range on heat flow. We show that, by increasing the interaction range, we may amplify
the thermal conductivity and even change the regime of heat transport. More importantly, considering a crucial problem of phononics, namely, the search of a
 suitable thermal diode, we investigate the range effects in some graded systems in which
thermal rectification is a ubiquitous phenomenon.
In such graded models, we show that long range interactions may significatively increase the rectification
power and may avoid its decay with the system size, thus solving relevant problems of the usual proposals of rectifiers. Our results indicate that graded materials are genuine
candidates for the actual fabrication of thermal diodes.

\end{abstract}


\maketitle

The invention of transistor used
to control the electric charge flow has led to the incredible
development of modern electronics. Now we observe the progress of phononics \cite{RMP}, the counterpart of electronics dedicated
to the manipulation and control of heat current. However, a very promising advance
is still dependent on the  development of one of its basic component: a realizable
thermal diode with a significative rectification.  A thermal diode, or rectifier, is a device in
which  heat flows preferably in one direction, i.e., the magnitude of the heat current changes if we invert the
device between two thermal baths.

 A model of thermal rectifier has been proposed some time ago\cite{Casati2}, and since then,  thermal rectification has been intensively investigated \cite{BLi1, BHu1, Prapid, Chang, BLiRapid, BLioutros, Psuf, WPC}, including experimental realizations \cite{Chang}. Unfortunately, the most recurrent proposals of thermal diode, which are based on the sequential coupling of two or three segments with different
anharmonic potentials,  are difficult to be experimentally implemented and their rectification power typically decays to zero when we increase the system size \cite{BHu1}. For these reasons, more and more efforts have
been devoted to the investigation of rectification in different models \cite{Prapid, Chang, BLiRapid, BLioutros}: for example, in graded systems (in which rectification is a ubiquitous phenomenon \cite{Psuf, WPC}), in carbon nano-structures, in systems composed of crystal and amorphous polymer, etc.  Graded materials are  inhomogeneous systems whose composition and/or
structure change gradually in space.
It is worth to stress that
such materials are abundant in nature, can
also be manufactured,  and have attracted great interest in many
areas \cite{Huang}, with works devoted to the study of their electric,
optical, mechanical and heat conduction properties.

In the search for mechanisms which may increase the rectification power, and/or avoid its rapid decay with the system size, the present work is devoted to a basic,
but somehow neglected problem: the effects of the interparticle interaction
range on the heat flow properties.  Given the enormous  mathematical difficulty of the usual models considered in the study of heat conduction
in solids, models which are described by systems with anharmonic on-site potentials and
harmonic nearest neighbor interparticle interactions, the investigation of systems with long range interactions seems
to be an exceedingly difficult task, but we will show that such an investigation
is feasible and provides useful results. We  still recall that the study of the role played by the range of the interaction
is a fundamental and ubiquitous problem in physics: it is responsible for different effects in classic and quantum
systems, in microscopic and macroscopic phenomena, in electronic transport, in equilibrium and nonequilibrium phase
transitions, etc\cite{refRic}.

We start the investigation by analyzing  homogeneous systems. We show that, by increasing the interaction range,
we can make bigger the thermal conductivity of a system with
normal heat transport, and we can even change the regime to anomalous transport.
For the case of asymmetric models, precisely, graded materials, we show that the introduction of interparticle interaction beyond nearest neighbor
sites may increase the rectification factor and still avoid its decay with the system size. That is, in a system with long range interactions, the rectification power may becomes hugely (thousand times) larger than that
observed in a similar system  with nearest neighbor interactions only. By long range we mean interactions with polynomial decay; and short range denotes those with exponential decay or compact support.
In other words, we show that the interaction range, due to these interesting effects, may be a key issue
in the search of materials with considerable thermal rectification, materials which are, as said, the basic ingredient for the building of thermal devices
of phononics such as thermal diodes and transistors.

We analyze recurrent microscopic models for heat conduction in solids, used since the pioneering work of
Debye, namely, chains of oscillators. And so, we believe that our results will be valid for real materials. Our formalism
includes harmonic and anharmonic chains with self-consistent stochastic reservoirs at each site, as well as
anharmonic chains with baths at the boundaries only. These models, for the specific case of nearest neighbor
interparticle interactions, obey the Fourier's law of heat conduction (see e.g. \cite{BLL, BK, Prapid}).

Let us introduce the models. For simplicity, we consider one-dimensional chains. We take $N$ oscillators with Hamiltonian
\begin{equation}\label{Hamiltonian}
H=\sum_{j=1}^{N}\left(\frac{p_j^2}{2m_j}+\frac{M_jq_j^2}{2}+\sum_{\ell \neq j}\frac{J_{j \ell}}{2}(q_j-q_{\ell})^2+\lambda \mathcal{ P}(q_j)\right)~,
\end{equation}
where $M_{j}\geq0$, $J_{\ell j}=J_{j\ell}$, $\mathcal{P}$ is the anharmonic on-site potential (and so, $\lambda=0$ for the specific
harmonic case). The dynamics is given by
\begin{equation}\label{dynamics}
dq_j=(p_j/m_j) dt, ~ dp_j=-\frac{\partial H}{\partial q_j} dt - \zeta_{j}p_j dt + \gamma_j^{1/2} dB_j,
\end{equation}
where $B_{j}$ are independent Wiener processes; $\zeta_{j}$ is the coupling between site $j$ and its reservoir (for the
models with baths only at the boundaries, $\zeta_{j}=0$ if $j$ is an inner site); and $\gamma_{j}= 2\zeta_{j}m_{j}T_{j}$, where
$T_{j}$ is the temperature of the $j$-th bath.

To study the energy current inside the system we define, as usual, the energy of the $j$-th oscillator as
\begin{equation}
H_j(q,p)=\frac{1}{2}\frac{p_j^2}{m_j}+U^{(1)}(q_j)+\frac{1}{2}\sum_{\ell > j}U^{(2)}(q_j-q_{\ell})~,
\end{equation}
where the expressions for $U^{(1)}$ and $U^{(2)}$, the local and the interparticle potentials, follow from Eq.$(\ref{Hamiltonian})$ and from $\sum_{j=1}^{N}H_{j} = H$. From the stochastic dynamics we get
\begin{eqnarray}
\left< \frac{dH_j}{dt}(t) \right>&=&\left< \mathcal{R}_j(t) \right> + \left< \mathfrak{F}_{\rightarrow j}-\mathfrak{F}_{j\rightarrow} \right>~,\\
\mathfrak{F}_{j\rightarrow}&=&\sum_{\ell>j}\nabla U^{(2)}(q_j-q_{\ell})\left(\frac{p_j}{2m_j}+\frac{p_{\ell}}{2m_{\ell}}\right) \nonumber \\
&=&\sum_{\ell>j}J_{j\ell}(q_j-q_{\ell})\left(\frac{p_j}{2m_j}+\frac{p_{\ell}}{2m_{\ell}}\right) ~\label{fluxo}\\
&=& \sum_{\ell>j} \mathfrak{F}_{j,\ell}~ \nonumber,
\end{eqnarray}
and a similar formula follows for $\mathfrak{F}_{\rightarrow j}$ (with the change between $j$ and $\ell$, and with the condition $\ell<j$).
In the equations above,  $\left< \cdot \right>$ denotes the expectation with respect to the noise distribution, and
$
\left<\mathcal{R}_j \right>=\zeta_j \left(T_j-\left<p_j^2\right>/m_j \right)
$
gives the energy flow between the $j$-th reservoir and the $j$-th site. In the steady state (as
$t \rightarrow \infty$), $\left<\mathcal{R}_{j}\right>$ always vanishes for the inner sites. Precisely, for an inner site $j$, we have
$\zeta_{j}=0$ for the case of a model with baths only at the ends; otherwise, the self-consistent condition is given by the choice of $T_{j}$ such that $\left<\mathcal{R}_{j}\right>=0$.
Such condition, i.e. the absence of mean heat flow between an inner site and its reservoir in the steady state, means that
inner reservoirs do not describe real thermal baths such as those reservoirs at the boundaries; they represent only some residual
interaction, some mechanism of phonon scattering not present in the deterministic potential. These systems with self-consistent inner stochastic reservoirs are old models \cite{Visscher}, recurrently studied \cite{BLL, BLLO}.

In the steady state we have $\left< dH_{i}(t)/dt \right>=0$, and so, the mean heat flow from site $j$ to site $\ell$, with $\ell >j$, is given by
$\mathcal{F}_{j\rightarrow} \equiv \left< \mathfrak{F}_{j\rightarrow} \right> = \sum_{\ell>j} \left< \mathfrak{F}_{j,\ell}\right> \equiv \sum_{\ell>j} \mathcal{F}_{j,\ell}$.

From Eq.$(\ref{fluxo})$, it follows that the heat flow in the system is given in terms of two-point functions $\left< q_{j}p_{\ell}\right>$. However, as well
known, the analysis of such two-point functions may require a very hard work. For the particular case of a chain with nearest neighbor interparticle
interaction, several works have been devoted to the problem. For systems with harmonic potentials and self-consistent
reservoirs, it is proved \cite{BLL, FFP} that
\begin{equation}\label{LocalFourier}
\mathcal{F}_{j,j+1}=\kappa_{j,j+1}(T_j-T_{j+1})~,
\end{equation}
where $\mathcal{F}_{j,j+1}$ is the flow from site $j$ to $j+1$; $\kappa_{j,j+1}$ does not depend on $T$: it is a function of $J_{j,j+1}$, of the particle mass, of the on-site harmonic potential
strength $M_{j}$, and of the coupling constant with the reservoirs. The approach and the integral formalism used to derive such results \cite{PF,FFP}
allow us to write a similar expression for $\mathcal{F}_{j,\ell}$ in the case of an interparticle potential with
interaction beyond next-neighbor sites, and so, we also have
\begin{equation}\label{LocalFourier2}
\mathcal{F}_{j,\ell}=\kappa_{j,\ell}(T_j-T_{\ell})~.
\end{equation}
In the case of  a system with nearest neighbor interparticle interactions, anharmonic on-site potential and self-consistent reservoirs, the approach involving an integral
representation for the heat flow \cite{Prapid, PFL} allows us to write $\mathcal{F}_{j,j+1}$ as Eq.$(\ref{LocalFourier})$, but with $\kappa_{j,j+1}$
depending on temperature. The same follows for the anharmonic, self-consistent chain with interactions beyond nearest neighbors, i.e., the heat flow is given by terms such as Eq.$(\ref{LocalFourier2})$, see Ref.\cite{Prapid}. For the anharmonic chain with baths only at the boundaries, the studies presented e.g. in Ref.\cite{BK}
lead to similar expressions for $\mathcal{F}_{j,j+1}$. To infer the behavior of $\mathcal{F}_{j,\ell}$, we
turn to Ref.\cite{PFL}, where a huge similarity between the anharmonic self-consistent chain and the anharmonic chain with
baths only at the ends is pointed out, at least for large anharmonicity.

We remark that expressions for $\kappa$ have been already precisely derived for some models. For example, for the homogeneous harmonic chain with self-consistent
reservoirs and weak nearest neighbor interaction ($|J_{j,j+1}|$ small), it is proved in Refs.\cite{PF,FFP} that
$$
\kappa_{j,j+1} \simeq \frac{J_{j,j+1}^{2}}{2\zeta M} .
$$
For the  self-consistent chain, with nearest neighbor interaction and anharmonic on-site potential given by $\lambda q^{4}$, we also have
$\kappa_{j,j+1}$ proportional to $J_{j,j+1}^{2}$   (see Ref.\cite{Prapid} for details). The relation $\kappa_{j,\ell} \varpropto J_{j,\ell}^{2}$ also follows for the anharmonic
 self-consistent chain with interaction beyond nearest neighbor sites: to see it, note that
 Eq.$(30)$ in the second work of Ref.\cite{Prapid} shows that the two-point function is proportional to the interparticle interaction; and recall that the heat flow is given by the product of the two-point function and the interparticle
interaction $J_{j,\ell}$ as described above in Eq.$(\ref{fluxo})$. We remark that the decay of the two-point function given by the decay of the interparticle interaction is also observed in other stochastic dynamics \cite{TPP}.
And, again, recalling Ref.\cite{PFL}, at least for highly anharmonic systems (large $\lambda$), we expect a
behavior for the heat flow in the chain with reservoirs at the boundaries similar to that observed in the self-consistent chain. In short, for many harmonic
and anharmonic chains (including those treated here), we have $\kappa_{j,\ell} \varpropto J_{j,\ell}^{2}$.

Thus, the models to be treated here,  given by chains of $N$ oscillators,  are such that, in the steady state, the energy current obeys the following equations
\begin{widetext}
\begin{eqnarray}
\lefteqn{\mathcal{F}=\kappa_{1,2}(T_1-T_2)+\kappa_{1,3}(T_1-T_3)+\ldots+\kappa_{1,N}(T_1-T_N)~,}\nonumber \\
 && \kappa_{1,j}(T_1-T_j)+\kappa_{2,j}(T_2-T_j)+\ldots+\kappa_{j-1,j}(T_{j-1}-T_j)=
\kappa_{j,j+1}(T_j-T_{j+1})+\ldots+\kappa_{j,N}(T_j-T_N).
\end{eqnarray}
\end{widetext}
The first equation means that all the energy which flows into the system comes from the first reservoir to the first site, and it is equal
to the energy that flows from the first site to the other ones. The second equation, which holds for $j=2, 3, \ldots, N-1$, means that all
the energy that comes from the previous sites to site $j$ is equal to the energy that leaves site $j$ to the following sites. For clearness, we
rewrite the system of equations above as
\begin{widetext}
\begin{equation}
\begin{array}{cccccccc}\label{sistema}
-\mathcal{F}&-\kappa_{1,2}T_2&-\kappa_{1,3}T_3 &\ldots&-\kappa_{1,N-2}T_{N-2}&-\kappa_{1,N-1}T_{N-1}&=&-\alpha_1T_1+\kappa_{1,N}T_N~,\\
0&+\alpha_2T_2&-\kappa_{2,3}T_3 &\ldots&-\kappa_{2,N-2}T_{N-2}&-\kappa_{2,N-1}T_{N-1}&=&\kappa_{2,1}T_1+\kappa_{2,N}T_N~,\\
\vdots&\vdots& & & &\vdots&=&\vdots\\
0&-\kappa_{j,2}T_2&\ldots&+\alpha_jT_j&\ldots&-\kappa_{j,N-1}T_{N-1}&=&\kappa_{j,1}T_1+\kappa_{j,N}T_N~,\\
\vdots&\vdots& & & &\vdots&=&\vdots\\
0&-\kappa_{N-1,2}T_2&-\kappa_{N-1,3}T_3&\ldots&-\kappa_{N-1,N-2}T_{N-2}&+\alpha_{N-1}T_{N-1}&=&\kappa_{N-1,1}T_1+\kappa_{N-1,N}T_N~,
\end{array}
\end{equation}
\end{widetext}
where $\alpha_{1} = \kappa_{12} + \kappa_{13} + \ldots + \kappa_{1N}$, $\alpha_{j} = \kappa_{j1} + \kappa_{j2} + \ldots + \kappa_{jN}$ (recall that $\kappa_{jj}=0$ and $\kappa_{\ell j} = \kappa_{j\ell}$).
For the simpler case of a thermal conductivity which does not depend on temperature,
given $T_{1}$ and $T_{N}$ we have  $N-1$ linear equations with $N-1$ variables: $\mathcal{F}, T_{2}, T_{3}, \ldots, T_{N-1}$.  It is also valid, in a first approximation, for a system submitted to a very small gradient
of temperature  (where we may write the thermal conductivity as a function of the average temperature instead of function of $T_{j}$, $T_{\ell}$).
For
the case of $\kappa$ given by a function
of the inner temperatures $T_{j}$, we do not have a simple linear system, and the solution is much more intricate.

A first scenario for the effects of the interaction range may be depicted by considering two extreme (opposite) cases in homogeneous models: first, a chain with nearest neighbor interaction; and, second,
a chain with a non-decaying $\kappa_{j\ell}$, i.e., with $\kappa_{12}= \ldots =\kappa_{1N}= \kappa_{j\ell} = \kappa$. Some comment is appropriate for the constant $\kappa$: it is certainly unphysical  and shall be considered only as an ``upper bound'' for the acceptable $\kappa$'s.

We describe below the mathematical solutions of Eqs.(\ref{sistema}) for the two extreme cases, but it is worth to note that such solutions, and the underlying physics, can also be derived by noting
the connection between Eqs.(\ref{sistema}) and circuits. Thus, we may also derive the solutions by using the Kirchhoff's theorem for circuits \cite{Referee}.

For nearest neighbor interactions and constant $\kappa$ (or $\kappa$ depending on the average temperature), our linear system becomes
\vskip0.2cm
\begin{tabular}{cccccc|c}
$\mathcal{F}$&$X_2$&$X_3$&$X_4$&$\ldots$&$X_{N-1}$& \\ \hline
$-1$&$-1$&$0$&$0$&$\ldots$&$0$&$-X_1$\\
$0$&$2$&$-1$&$0$&$\ldots$&$0$&$X_1$\\
$0$&$-1$&$2$&$-1$&$\ldots$&$0$&$0$\\
& &$\ddots$&$\ddots$&$\ddots$& &$\vdots$\\
$0$&$0$&$\ldots$&$-1$&$2$&$-1$&$0$\\
$0$&$0$&$\ldots$&$0$&$-1$&$2$&$X_N$\\
\end{tabular}
\vskip0.2cm
\noindent where $X_{j}= \kappa T_{j}$, $j=1,2, \ldots, N$. The last column, of the independent terms, involves $X_{1}$ and $X_{N}$. By using Crammer's rule, we have
$\mathcal{F} = \Delta_{\mathcal{F}}/\Delta$, where $\Delta$ is the determinant of the coefficient matrix, and $\Delta_{\mathcal{F}}$ is the determinant of the
matrix obtained from the coefficient matrix with the replacement of the first column by the column of the independent terms. We have $\Delta_{\mathcal{F}} =
-X_{1}D_{n-2} + X_{1}D_{n-3} + X_{N}$, where $D_{n-j}$ is the determinant of the Laplacian matrix with $n-j$ lines and columns (the Laplacian matrix is
given by the coefficient matrix above without the first line and the first column). It is easy to prove (e.g. by induction) that $D_{n} = n+1$. Then, we have
\begin{equation} \label{Fourier}
\mathcal{F}=\frac{\Delta_{\mathcal{F}}}{\Delta}=\frac{X_N-X_1}{-(N-1)}=\frac{\kappa(T_1-T_N)}{N-1}~.
\end{equation}
That is a very well known result: for the anharmonic chain or harmonic chain with self-consistent reservoirs, Fourier's law holds, i.e. the heat flow decreases with $N$. The underlying
physics is clear: the chain with nearest couplings may be understood as segments in series, in which each segment obeys a local Fourier's law (\ref{LocalFourier}). Hence, Fourier's law also holds in the whole chain, and so, the heat flow decays with the system size.

Let us turn to the other extreme case: $\kappa_{j\ell} =\kappa$, with $\kappa$ constant or depending on the average temperature. The linear system for
the heat flow and inner temperatures becomes
\vskip0.2cm
\begin{tabular}{cccccc|c}
$\mathcal{F}$&$X_2$&$X_3$&$X_4$&$\ldots$&$X_{N-1}$& \\ \hline
$-1$&$-1$&$-1$&$-1$&$\ldots$&$-1$&$X_N-(N-1)X_1$\\
$0$&$(N-1)$&$-1$&$-1$&$\ldots$&$-1$&$X_N+X_1$\\
$0$&$-1$&$(N-1)$&$-1$&$\ldots$&$-1$&$X_N+X_1$\\
 &$\vdots$& &$\vdots$& &$\vdots$&$\vdots$\\
$0$&$-1$&$-1$&$-1$&$\ldots$&$(N-1)$&$X_N+X_1$\\
\end{tabular}
\vskip0.2cm
\noindent where $X_{j}=\kappa T_{j}$, $j=1, \ldots, N$. First, let us compute the determinant $\mathcal{D}$ of the matrix where the elements are $1$ outside the main diagonal, and are $\alpha$ in the
diagonal. For a matrix with two lines and two columns we have: $\mathcal{D}_{2} = \alpha^{2}-1 = (\alpha-1)(\alpha+1)$. And $\mathcal{D}_{3} = (\alpha-1)(\alpha-1)(\alpha+2)$.
By induction we prove that $\mathcal{D}_{n} = (\alpha-1)^{n-1}(\alpha+n-1)$.

Turning to the heat flow, as before, we have $\mathcal{F} = \Delta_{\mathcal{F}}/\Delta_{c}$, where the determinant of the coefficient matrix is $\Delta_{c} = -1\cdot(-1)^{N-2}\mathcal{D}_{N-2}$,
with $\mathcal{D}_{N-2}$ computed with $\alpha=-(N-1)$. With some algebra, we get $\Delta_{c} = -2N^{N-3}$.  $\Delta_{\mathcal{F}}$ is the determinant of the coefficient matrix  with the first
column replaced by the column of independent terms. By developing the determinant in terms of the new first column,  we obtain for $\Delta_{\mathcal{F}}$
\begin{eqnarray}
\Delta_{\mathcal{F}}&=&[X_{N}-(N-1)X_1](-1)^{N-2}\mathcal{D}_{N-2}  \\
 & & -(X_N+X_1)(N-2)(-1)^{N}\left[\frac{\alpha\mathcal{D}_{N-2}-\mathcal{D}_{N-1}}{N-2}\right]~. \nonumber
\end{eqnarray}
With some algebraic manipulation, we get $\Delta_{\mathcal{F}} = N^{N-2}(X_{N}-X_{1})$, and so
\begin{equation}\label{NFourier}
 \mathcal{F}=\frac{N \kappa}{2}(T_1-T_N)~.
\end{equation}
That is, now the heat flow increases with $N$. To understand the physics behind such result, note that, for the fully coupled system, the
number of sites gives the possible channels (ways) for the heat current; hence, the flow is expected to enlarge as we increase the number of channels, i.e.,
the system size.

From the study of these previous limiting case, the effects of changes in the
interparticle interaction range on heat flow
are already quite clear: by increasing the interaction range we expect to make bigger the thermal conductivity and, for
long range, we even expect to change the regime to anomalous transport.

The detailed analytical investigation of the system of equations
$(\ref{sistema})$ for the case of $\kappa_{j\ell}$ decaying as a function
of $|j-\ell|$ and/or $\kappa_{j\ell}$ as a function of the temperatures $T_{j}$ and $T_{\ell}$ requires a very difficult work.
Hence, to analyze such cases we make use of numerical techniques \cite{NT}.

First we study systems submitted to a very small gradient of temperature, i.e., cases in which $\kappa$ may be considered as
a function of the average temperature. For an exponential decay $\kappa_{j\ell} = 1/2^{|j-\ell|}$, numerical computations show that the system still obeys Fourier's law
and the thermal conductivity increases as compared to system with nearest neighbor interaction. See Fig.1(a).
Considering a non-integrable polynomial decay, such as $1/|j-\ell|^{\gamma}$ with $\gamma \leq 1$, the computation gives a
 behavior which is similar to the extreme case with equal $\kappa$'s, that is,
the heat flows grows up as we increase $N$: as checked out for several $\gamma$, it seems that it grows like $\int_{1}^{N} (1/x^{\gamma})dx$, i.e., as $\ln N$ for $k_{j\ell}\sim 1/|j-\ell|$, etc.
See Fig.1(b).

\begin{figure}[h]
\includegraphics[scale=1.5,height=3cm,width=9cm]{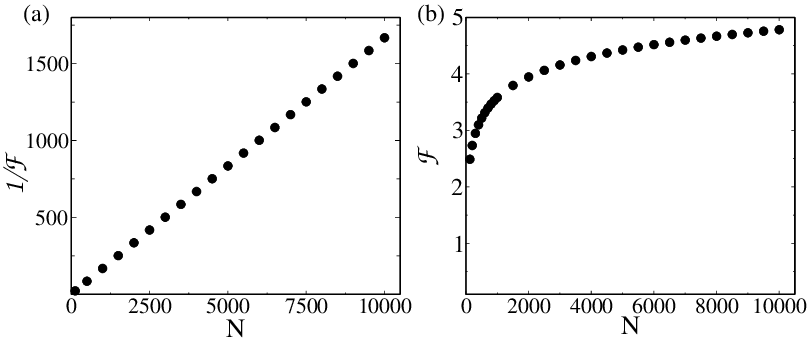}
\caption{Heat flow ${\cal F}$ versus number of sites for a homogeneous chain, where $T_{1}=2$, $T_{N}=1$,  and: (a)  $\kappa_{j\ell}$ decays  as $1/2^{|j-\ell|}$; note that
Fourier's law still holds with thermal conductivity $\kappa\simeq 6$, which is bigger than the thermal conductivity $\kappa=1/2$ of the chain with nearest neighbor interaction. (b)  $\kappa_{j\ell}$ decays as $1/|j-\ell|$; here the heat flow grows with $\ln N$.}
\end{figure}

Now, we turn to the main problem: we consider asymmetric chains and investigate the effects of long range interactions on thermal rectification.

As well known \cite{Prapid, Psuf, WPC, AP}, the dependence of thermal conductivity on local temperatures plays a crucial role in thermal rectification. Hence, we investigate systems in which $\kappa$ involves functions of temperature. Mimicking the behavior observed in graded chains of oscillators with anharmonic on-site potentials \cite{Prapid}, now we take $\kappa_{j\ell}$ as $g(|j-\ell|)/(c_{j}T_{j}^{\beta} + c_{\ell}T^{\beta}_{\ell})$, where the exponent $\beta$ gives the intricate nonlinear dependence of $\kappa$ on $T$; $g$ gives the distance decay, e.g. $g(|j-\ell|)\sim 1/|j-\ell|^{\gamma}$; and  $c_{j}$ is a term which depends on local graded parameters, e.g. on the particle mass (the asymmetry in the chain will be given by different terms $c_{j}$).

For such $\kappa$, even a numerical study becomes intricate. Thus, to follow with the computations on the heat flow, we restrict the analysis to a system submitted to a small temperature gradient:  we take a chain in which
$T_{1} = T + a_{1}\varepsilon$ and $T_{N}= T + a_{N}\varepsilon$, for some small $\varepsilon$. Consequently, the inner temperatures will be given in terms of $T$ and $\varepsilon$: up to second order in $\varepsilon$, we will have $T_{j} = T + a_{j}\varepsilon + b_{j}\varepsilon^{2}$ ($a_{j}$ and $b_{j}$ to be determined by solving $(\ref{sistema})$). Then, by expanding $\kappa_{j\ell}$ up to second order in $\varepsilon$, and solving the system
of equations $(\ref{sistema})$, we can determine all inner temperatures $T_{j}$ and $\mathcal{F} = \mathcal{F}_{1}\varepsilon + \mathcal{F}_{2}\varepsilon^{2}$ in terms of the
temperatures at the boundaries. Note that such procedure gives us two systems of equations: one for $\varepsilon$ and another for $\varepsilon^{2}$. To search for rectification, besides $\mathcal{F}$, we still
have to study the heat flow in the system as we invert the temperatures at the boundaries. That is, we also need to obtain the heat flow $\mathcal{F}'$ in
the system with temperatures $T'_{j}=T+a'_j\varepsilon + b'_j\varepsilon^2$, where
$T'_{1}=T_{N}$ and $T'_{N}=T_{1}$.

To start the investigation, we take a graded chain in which $c_{j}$ (related, e.g., to the particle mass) grows linearly with $j$, and $k_{j\ell}$ decays as $1/|j-\ell|^{1.1}$. We take this exaggerated
slow decay in order to make more transparent the effects. Then, we compute the rectification factor for the systems, which is defined as the difference
between the magnitude of the direct and reverse heat flow divided by the smaller one (the reverse flow is that obtained by inverting the temperatures at
 the boundaries). The rectification factor for a chain with linear graded mass distribution and $\kappa$ with polynomial decay is depicted in Fig.2(a). The ratio between the rectification factor of
 the case in which $\kappa$ has a polynomial decay and the case in which $\kappa$ is nearest neighbor is depicted in Fig.2(b). These results make transparent the considerable effect of the range interaction on  the rectification power.

 Trying to fix the other recurrent problem of diodes,  namely, the decay of rectification  with the system size, we turn now to more asymmetric systems. We take chains with exponential mass distribution, and compare the cases with long range polynomial decay and nearest neighbor interactions. Again, the effect is clear: the system with long
 range interaction has a much bigger rectification power. See Fig.2(d). Moreover, in contrast to the nearest neighbor interaction, the rectification power does not decay with the system size. See Fig.2(c).

\begin{figure}[h]
\includegraphics[scale=1.1,height=6cm,width=9cm]{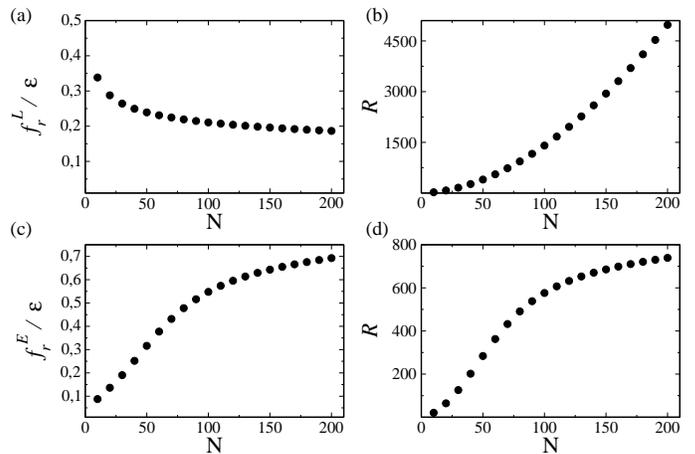}
\caption{Rectification factor $f_r^i$, in multiples of $\varepsilon\simeq |T_{1}-T_{N}|$, versus number of sites. Here: $T=1, a_{1}=2, a_{N}=1, \beta=1.5$. In (a), index $i=L$ denotes a linear graded mass distribution, $c_j=j$. In (c), $i=E$ denotes an exponential graded mass distribution, $c_j=\exp(-\delta j)$, with $\delta=0.05$. For both cases,  $\kappa_{j,\ell}=c/|j-\ell|^{1.1}$, $c=100$. The ratio $R$ between the rectification factors for long range and nearest neighbor cases versus number of sites is depicted, in (b) and (d), for linear and exponential graded mass distribution, respectively.}
\end{figure}

We remark that the rectification factors computed above are multiplies of  $\varepsilon$, i.e., of the temperature difference  $|T_{1}-T_{N}|$, which is here, due to technical difficulties, a small amount. However, as shown by theoretical analysis \cite{AP} and simulations \cite{WPC} in
previous works with nearest neighbor interactions,
 the dependence of the rectification factor on
 $|T_{1}-T_{N}|$ as described here is certainly valid beyond the regime of small differences. In other words, for larger differences of temperature we certainly obtain bigger (and significative)
 rectification factors.

In all the cases considered in the study of rectification above, the heat flow is given by, up to $\mathcal{O}(\varepsilon^{2})$,
$\mathcal{F} = \mathcal{F}_{1}\varepsilon + \mathcal{F}_{2}\varepsilon^{2}$. For small gradients of temperature, the rectification appears only in
$\mathcal{F}_{2}$, leading to a  $\mathcal{O}(\varepsilon)$ rectification factor, which is defined, as already said,  as the difference between the direct and the reversed flows divided by the smaller one. In Fig.3, the heat flow term $\mathcal{F}_{2}$ and the reversed one $\mathcal{F}'_{2}$ are depicted
for the cases treated in Fig.2. The total direct and reversed heat flows $\mathcal{F}$ and $\mathcal{F}'$, not plotted in the graphics,  increase with the system size $N$ for the exponential mass distribution, and decay for
the linear case.

\begin{figure}[h]
\includegraphics[scale=1.7,height=3.5cm,width=9cm]{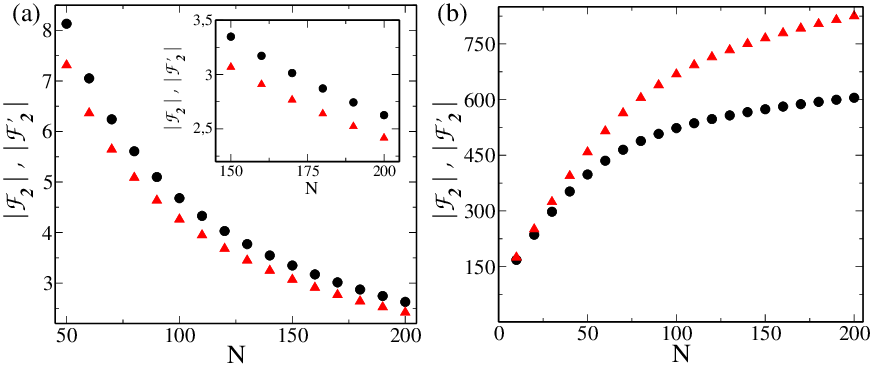}
\caption{(Color online) Modulus of direct heat flow $|\mathcal{F}_{2}|$ (triangles) and reversed one $|\mathcal{F}'_{2}|$ (circles) versus number of sites for inhomogeneous chains. Here: $T=1, a_{1}=2, a_{N}=1, \beta=1.5$.  A linear graded mass distribution, $c_j=j$, and an exponential one, $c_j=\exp(-\delta j)$, with $\delta=0.05$, are assumed in (a) and (b), respectively.  For both cases,  $\kappa_{j,\ell}=c/|j-\ell|^{1.1}$, $c=100$.}
\end{figure}

Further investigation with more detailed results by means of computer simulations are highly desirable, but we leave such task for the experts.

We still need to make some comments regarding the connection between our results and real materials. First of all, we recall, as repeatedly said throughout the paper, that graded systems are found in nature and can be also manufactured. Moreover, nowadays several materials, such as nanomagnets, can be fabricated and even manipulated  with the lithographical insertion of different types of pointlike magnetic impurities
in order to present certain properties. A good example is given by magnets of nanodisks of Permalloy \cite{Bis}, material in which the interparticle interaction presents a polynomial decay: $1/r_{ij}^{3}$. In short, graded chains and systems with  interactions polynomially decaying are not only theoretical models. However, some conditions assumed in our analysis, e.g. the exponential graded mass distribution and the slow polynomial decay of the
interaction, may be extremely difficult to be approached in real materials. We have used such mass distribution
 in order to obtain a very asymmetric chain. A more realistic description, still giving a very asymmetric chain, may be obtained by taking a more acceptable graded mass distribution together with graded interparticle potentials, with graded on-site anharmonic potentials, etc., i.e., with other asymmetric characteristics.
Anyway, we understand that a study with exaggerated conditions is still useful to make
transparent, to amplify effects which will survive,  without such intensity, in a more realistic situation.

In summary, in the present work, we investigate the effects of range interaction on the heat flow. We show that interactions beyond nearest neighbor sites may
increase the thermal conductivity and even change the transport regime, properties of practical interest. More importantly, addressing a crucial problem of phononics, we show that long range interactions may considerably
increase the rectification power and may avoid its decay with the system size, problems of the usual proposals of rectifiers. In particular, we show that such phenomenon occurs in graded systems, realizable materials in which thermal rectification ubiquitously holds.
In short, our results indicate that graded materials are genuine candidates for the actual fabrication of thermal diodes.

We are in debt to an anonymous referee, who called our attention to the connection between some of our results and the Kirchhoff's theorem
for circuits \cite{Referee}.
We thank R. Sardenberg, M. Matos, M. C. Aguiar and B. V. Costa for the help with numerical programs. This work was partially supported by CNPq (Brazil).


\begin{thebibliography}{32}
\bibitem{Referee} An anonymous referee called our attention to the fact that  Eqs.(\ref{sistema}), describing  the heat flow in the
system, correspond to an effective eletric (or heat) circuit. Hence, our results given by Eqs.(\ref{Fourier}, \ref{NFourier}) can be
directly derived from the Kirchhoff's theorem for circuits. Precisely, the case of nearest neighbor coupling corresponds to the circuit in
series, and so, its conductivity is $\kappa/(N-1)$, decreasing with size. The case of fully connections corresponds to the circuit in parallel (that is not so easy to see),
so that its conductance is $\kappa + (N-2)\kappa/2 = N\kappa/2$, increasing with size. By considering such relation with circuits, the underlying
physics is obvious: in the circuit in series, the resistance increases with $N$, and so, the conductance decays; the opposite behavior follows for
the circuit in parallel.
\bibitem{NT} Concerning the numerical analysis, we repeat the computation three times by using
 Mathematica, Mathcad and a program in Fortran.
\end{thebibliography}
\end{document}